# Interfacial Properties of Monolayer and Bilayer MoS$_2$ Contacts with Metals: Beyond the Energy Band Calculations

Hongxia Zhong,[1,†] Ruge Quhe,[1,3,†] Yangyang Wang,[1,5] Zeyuan Ni,[1] Meng Ye,[1] Zhigang Song,[1] Yuanyuan Pan,[1] Jinbo Yang,[1,2] Li Yang,[4] Junjie Shi,[1,*] and Jing Lu[1,2,*]

[1]State Key Laboratory for Mesoscopic Physics and Department of Physics, Peking University, Beijing 100871, P. R. China

[2]Collaborative Innovation Center of Quantum Matter, Beijing 100871, P. R. China

[3]State Key Laboratory of Information Photonics and Optical Communications, Beijing University of Posts and Telecommunications & School of Science, Beijing 100876, China

[4]Department of Physics, Washington University in St. Louis, St. Louis, Missouri 63130, USA

[5]Department of Nuclear Science and Engineering and Department of Materials Science and Engineering, Massachusetts Institute of Technology, Cambridge, Massachusetts 02139, USA

[†]These authors contributed equally to this work.

Email: jjshi@pku.edu.cn; jinglu@pku.edu.cn

## ABSTRACT

Although many prototype devices based on two-dimensional (2D) MoS$_2$ have been fabricated and wafer scale growth of 2D MoS$_2$ has been realized, the fundamental nature of 2D MoS$_2$-metal contacts has not been well understood yet. We provide a comprehensive ab initio study of the interfacial properties of a series of monolayer (ML) and bilayer (BL) MoS$_2$-metal contacts (metal = Sc, Ti, Ag, Pt, Ni, and Au). A comparison between the calculated and observed Schottky barrier heights (SBHs) suggests that many-electron effects are strongly suppressed in channel 2D MoS$_2$ due to a charge transfer. The extensively adopted energy band calculation scheme fails to reproduce the observed SBHs in 2D MoS$_2$-Sc interface. By contrast, an ab initio quantum transport device simulation better reproduces the observed SBH in the two types of contacts and highlights the importance of a higher level theoretical approach beyond the energy band calculation in the interface study. BL MoS$_2$-metal contacts have a reduced SBH than ML MoS$_2$-metal contacts due to the interlayer coupling and thus have a higher electron injection efficiency.

**Subject Areas:** Computational Physics, Electronics, Nanophysics

# I. INTRODUCTION

Owing to their excellent properties, two-dimensional (2D) molybdenum disulfide $MoS_2$ has attracted much recent attention [1-6]. A variety of prototype devices based on 2D $MoS_2$ have been fabricated, such as field-effect transistors (FETs) [7-9], inverters [10], fully integrated circuits [11], sensors [12], photoelectronic devices [13], phototransistors [14,15], spintronic devices [16], and valleytronic devices [17,18]. Very recently, wafer-scale high performance 2D $MoS_2$ FETs have been fabricated in batch mode, paving the way towards atomically thin integrated circuitry [19]. Among 2D $MoS_2$, monolayer (ML) and bilayer (BL) $MoS_2$ attract the most attention [2,3,5,6-18]. They show quite interesting differences and make up a pair of complementary materials: (1) ML $MoS_2$ has a larger direct band gap, while BL $MoS_2$ possesses a smaller indirect band gap due to the strong interlay coupling. Correspondingly, photoluminescence is dramatically enhanced in ML $MoS_2$ [6,20]. (2) ML $MoS_2$ is inversion asymmetric and serves as an ideal valley Hall insulator (VHI) [1]. By contrast, inversion symmetric BL $MoS_2$ is not a VHI, but it can be transformed into a VHI with a tunable valley magnetic moment by a vertical electric field, which destroys the inversion symmetry [5]. (3) Zeeman-like spin splitting is nearly intact by a vertical electric field in ML $MoS_2$ but it becomes tunable in BL $MoS_2$ because top and bottom $MoS_2$ feel different electric potentials [16].

In a real device, semiconducting 2D $MoS_2$ needs a contact with metal electrodes, and a Schotty barrier is often formed in semiconductor-metal interface, which impedes the carrier transport. The formation of low-resistance metal contacts is the biggest challenge that masks the intrinsic exceptional electronic properties of 2D $MoS_2$, and many efforts have been made to study 2D $MoS_2$-metal contact so as to reduce the Schottky barrier height (SBH) [21-23]. The SBH of a 2D $MoS_2$-metal contact depends on the work function of metal and the layer number of $MoS_2$. Lower work function metal and more $MoS_2$ layer number favor a smaller SBH. For example, there is a significant SBH between Ti and ML $MoS_2$ [24,25]；by contrast, Ti forms an Ohmic contact with BL $MoS_2$ at room temperature and a Schottky contact with a small SBH of ~ 0.065 eV at a low temperature [11,26]. Although there are several energy band calculations based on single particle density functional theory (DFT) to examine ML $MoS_2$-metal interfaces [22,23,25,27,28], a comprehensive energy band calculation for BL

MoS$_2$-metal interfaces is still lacking at present.

There are two open issues concerning this validity of the DFT energy band approach to treat the SBH of a transistor. Because the SBH at the metal-semiconductor interfaces depends on the difference between the Fermi level ($E_f$) of the metal and the band edge positions of the semiconductor, the band edge positions of the semiconductor must be accurately determined [29,30]. It is well known that the DFT fails to do so because it is a single-electron theory. From a theoretical point of view, the accurate band edge positions should be the quasiparticle energy, which can be obtained from first-principles many-electron Green function approach within the *GW* approximation, where electron-electron correlation effects are treated properly [13,31-38]. The first open issue concerning the DFT energy band scheme to evaluate SBH is whether the many-electron effects should be included.

The second open issue is the way of the energy band calculation in treating the SBH of a transistor. There are two possible interfaces to form Schottky barrier in a MoS$_2$ transistor [23,39]: one is the source/drain interface (B) between the contacted MoS$_2$ and the metal surface in the vertical direction (see Fig. 7(c)) if the interaction between MoS$_2$ and metal is weak, and the other is source/drain-channel (D) interface between the contacted MoS$_2$ and channel MoS$_2$ in the lateral direction (see Fig. 7(c)) if a metallization has taken place between MoS$_2$ and metal. The energy band calculation treats the source and the channel independently and ignores the coupling between the source and the channel, which may lead to the Fermi level pinning and change the SBH.

In this Article, we provide a theoretical study of the interfacial properties of ML and BL MoS$_2$ on several commonly used metals (Sc, Ti, Ag, Pt, Ni, and Au) [8,21] at different levels. A comparison between the observed and calculated SBH in ML and BL MoS$_2$-Ti interfaces suggests that *GW* correction to the band edge positions of 2D MoS$_2$ is strongly depressed in a device because of a charge transfer. More importantly, we find that the energy band calculation is unable to reproduce the observed SBHs in 2D MoS$_2$-Sc and -Pt interfaces. This failure prompts us to perform direct an ab initio quantum transport device simulation, and we find the SBHs in 2D MoS$_2$-Sc and -Pt interfaces can be better reproduced in latter calculation. SBH is found to be reduced from ML-metal interfaces to BL MoS$_2$-metal interfaces in different level calculations.

## II. METHODOLOGY

We use six layers of metal atoms (Ni, Ag, Pt, and Au in (111) orientation and Sc and Ti in (0001) orientation) to model the metal surface and construct a supercell with ML and BL MoS$_2$ adsorbed on one side of the metal surface. BL MoS$_2$ takes AB stacking mode (with a $D_{3d}$ point group symmetry) in our model. The calculated in-plane lattice constant $a$ = 3.166 Å, which is in good agreement with the experimental value 3.160 Å [40]. The MoS$_2$ 1 × 1 unit cell is adjusted to the 1 × 1 unit cells of Sc and Ti(0001) faces, and the MoS$_2$ $\sqrt{3}\times\sqrt{3}$ unit cell is adjusted to 2 × 2 unit cells [27]. The lattice constant mismatches with respect to that of MoS$_2$ are 1.2 ~ 9.1%. A vacuum buffer space of at least 15 Å is set to ensure decoupling between neighboring slabs. MoS$_2$ mainly interacts with the topmost three layers metal atoms [22], so cell shape and the bottom three layers of metal atoms are fixed.

The geometry optimization and electronic properties of the periodic structures are performed using the projector augmented wave (PAW) method implemented in the Vienna *ab initio* simulation package (VASP) code [41,42]. The generalized gradient approximation (GGA) functional to the exchange-correction functional, of the Perdew–Wang 91 (PW91) form [43] with vdW corrections [44], and the PAW pseudopotential are adopted [42]. The cut off energy is set to 500 eV after convergence tests. An equivalent Monkhorst-Pack $k$-points grid [45] of 25 × 25 × 1 for a MoS$_2$ unit cell is chosen for supercell relaxations and 30 × 30 × 1 for property calculations. In our current calculations, the total energy is converged to less than $10^{-5}$ eV. Dipole corrections in the $z$ direction are used in all calculations. The maximum force is less than 0.02 eV/Å during optimization.

Transport properties of the gated two-probe model is established by the DFT coupled with the nonequilibrium Green's function (NEGF) method, as implemented in the ATK 11.8 package [46, 47]. We employ the single-zeta plus polarization (SZP) basis set during the device simulation. A test using higher double-zeta plus polarization (DZP) basis set is also performed. In consistent with previous DFT calculations, GGA of PW91 form to the exchange-correlation functional is used through the device simulations. The Monkhorst-Pack $k$-point meshes for the central region and electrodes are sampled with 1×50×1 and 50×50×1 separately. The temperature is set to 300 K. The Neumann condition is used on the boundaries of the direction vertical to the MoS$_2$ plane. On the surfaces connecting the electrodes and the central region, we employ Dirichlet boundary condition to ensure the charge neutrality in the source and the drain region. The transmission coefficient T($E$) is given by T($E$) = $G(E)\Gamma^L(E)G^\dagger(E)\Gamma^R(E)$, where $G(E)$ and $G^\dagger(E)$ are the retarded and advanced Green functions, and the broadening function $\Gamma^{L/R}(E)$ describes the level broadening due to left/right electrode and is obtained from the electrode self-energies $\Gamma_{L/R}(E) = i(\Sigma_{L/R} - \Sigma^\dagger_{L/R}))$. The electrode self-energies can be viewed as an effective Hamiltonian describing the interaction between device and lead.

## III. RESULTS AND DISCUSSION
### A. Geometry and stability of ML and BL MoS$_2$-metal interfaces

After optimizing the structures from 6 initial configurations in an interface with 1×1 MoS$_2$ unit cell and 3 initial configurations in an interface with $\sqrt{3}\times\sqrt{3}$ MoS$_2$ unit cell, we obtain the most stable configurations of the ML MoS$_2$-metal interfaces, as shown in Fig. 1. The initial configurations of BL MoS$_2$-metal interfaces are constructed on the basis of the most stable ML MoS$_2$-metal interfaces. On Sc(0001), the Mo atoms in the primitive cell sit above the top metal atom, and the S atoms sit above the second layer metal atom; While on Ti(0001), the Mo atoms in the primitive cell still sit above the top metal atom, but the S atoms sit above the centers of triangles. On Ni and Pt(111), the three Mo atoms in the supercell sit above the fcc hollow, hcp hollow, and top sites, respectively. In the cases of Ag and Au(111), the Mo atoms are all above the centers of the triangles formed by the fcc, hcp, and top sites. The calculated key parameters of ML and BL MoS$_2$-metal interfaces studied in this work are summarized in Table 1. When ML and BL MoS$_2$ are in contact with metals, the equilibrium distances $d_{S-M}$ range from 1.557 ~ 3.405 Å, increasing in the order of Ti < Sc < Ni < Pt < Ag < Au. The binding energies $E_b$ have a reversal order, i.e., Ti > Sc > Ni > Pt > Ag > Au, since a smaller $d_{S-M}$ generally causes a larger binding energy. Ti and Sc have a strong adhesion with ML/BL MoS$_2$ ($E_b$ = 1.181 ~ 1.848 eV per surface sulfur atom), Ni, Pt, and Ag have a medium adhesion ($E_b$ = 0.503 ~ 0.830 eV per surface sulfur atom), and Au has a weak adhesion ($E_b$ = 0.307 ~ 0.354 eV per surface sulfur atom). ML and BL MoS$_2$-metal contacts nearly share the same $d_{S-M}$ and $E_b$. The binding of MoS$_2$ to metal surfaces [22,27] is considerably stronger than that of graphene to metal surfaces, with the binding energy of 0.027 ~ 0.327 eV per carbon atom [48]. Such a difference is reasonable because MoS$_2$ is chemically more reactive than graphene. We note that previous DFT calculations indicate that the $E_b$ of ML MoS$_2$ on Ir, Pd, and Ru surfaces ranges from 0.62 ~ 0.82 eV per surface sulfur atom [27].

### B. Electronic structure of ML and BL MoS$_2$-metal interfaces

The electronic structures of free-standing ML MoS$_2$ and the interfacial systems are presented in Fig. 2. Free-standing ML MoS$_2$ has a direct band gap of 1.68 eV, consistent with the reported PBE value of 1.67 eV [49]. The band structures of ML MoS$_2$-metal contacts are classified into three categories in terms of the hybridization degree of ML MoS$_2$ on metals. The band structure of ML MoS$_2$ is identifiable clearly for MoS$_2$ on Au surface (weak hybridization), as a result of weak charge-transfer interaction and dispersion interaction between ML MoS$_2$ and Au surface. The band structure of ML MoS$_2$ is destroyed seriously (strong hybridization) by Sc and Ti surfaces and destroyed but still identifiable (medium hybridization) by Ni, Pt, and Ag surfaces, because the outmost electrons of the five metals except Ag are $d$ electrons, which strongly hybridize with the states near the Fermi level $E_f$ of ML MoS$_2$. For the sake of comparison, the electronic structures of free-standing BL MoS$_2$ and BL MoS$_2$-metal interfaces are also shown in Fig. 3, with a smaller indirect band gap of 1.46 eV for free-standing BL MoS$_2$. The band hybridization degree is similar from ML to BL MoS$_2$ and can be also divisible into the same three categories. The hybridization degree of ML/BL MoS$_2$ on metals is consistent with its binding energy: The higher the binding energy is, the higher the hybridization degree is.

In order to have a deep understanding for the hybridization in Figs. 2 and 3, we further calculate the partial density of states (PDOS) on Mo and S orbitals for ML and BL MoS$_2$-metal contacts as shown in Figs. 4 and 5. Upon making a contact with Sc and Ti, a large amount of Mo and S states are extended into the original band gap of ML/BL MoS$_2$ due to metallization. In the MoS$_2$-Sc system, the contribution of S 3$sp$ and Mo 4$d_{xy}$ states dominate $E_f$, which is associated with a strong S-Sc mixing. $E_f$ is dominated by Mo 4$d_{xy}$ states, with the other states playing a minor role in the MoS$_2$-Ti system. Mo and S states also appear in the original MoS$_2$ band gap due to orbital overlap between MoS$_2$ and metal. There is no Mo and S state in the original MoS$_2$ band gap in MoS$_2$-Au system, indicating that MoS$_2$ preserves the semiconducting nature on Au surface.

Large charge carrier density at the source/drain interface B indicates a strong overlap of electron orbitals and sufficient injection of electron into the $MoS_2$ layer [22]. The electron densities averaged in planes parallel to the interface $\rho_l$ of the investigated six ML $MoS_2$-metal contacts are displayed in Fig. 6. We can see from Fig. 6 that $\rho_l$ at the strong bonding interfaces (Sc, Ti, Ni, and Pt) is higher than that at the weak bonding interfaces (Ag and Au), a difference compared with the PDOS analysis in Fig. 4. This difference implies that the chemisorption interface have a larger possibility to achieve a lower contact resistance.

### C. Many-electron effects

The accurate SBH at a metal-semiconductor interface depends on the absolute band-edge positions of the semiconductor. Because the DFT method seriously underestimates the band gap of a semiconductor, the inclusion of the *GW* correction is also necessary to obtain a correct band gap and absolute band-edge positions of a freestanding (or undoped) semiconductor. If the band gap center (BGC) or Fermi level $E_f$ or work function and the *GW* corrected band gap ($E_g^{GW}$) of the semiconductor are available, the absolute energies at the conduction band maximum (CBM) and the valence band minimum (VBM) can be obtained via the relation:

$$E_{CBM}^{GW} = E_f + \frac{1}{2} E_g^{GW} \qquad (1)$$

$$E_{VBM}^{GW} = E_f - \frac{1}{2} E_g^{GW} \qquad (2)$$

Unfortunately, in many cases, the BGC of a semiconductor is unavailable. A common solution [29,30] is to assume that the BGC at the DFT level is unchanged after the *GW* correction (named *GW*-BGC approximation). Figs. 7(a) and (b) illustrate the *GW* correction to the absolute band positions for freestanding ML and BL $MoS_2$ in BGC approximation. Based on the *GW*-BGC scheme, the calculated ionization potential (IP = 5.45 eV) and electron affinity ($\chi$ = 4.22 eV) of bulk $MoS_2$, compared with values of 5.33 and 4.45 eV at the DFT level [29], are in good agreement with the experimental values (IP = 5.47 ± 0.15 eV and $\chi$ = 4.07 ± 0.35 eV) [50]. Actually, Yang *et al.* [51] found that the absolute band-edge energies for ML dichalcogenides given by the direct *GW* method and the *GW*-BGC scheme are quite similar. Therefore, the *GW*-BGC approximation is a good approximation for our studied

MoS$_2$ systems.

The *GW* corrections to the band gap of free-standing ML ($E_g^{GW}$ = 2.84 eV) and BL ($E_g^{GW}$ = 1.82 eV) MoS$_2$ are available [36,37]. In our calculations for the SBH at the vertical direction, we take the *GW*-BGC approximation to determine the absolute band edge positions. In our calculations for the SBH at the lateral direction, we determine the *GW*-corrected absolute band position by taking the experimental work function (5.25 eV [52]) of free-standing BL MoS$_2$ (namely channel BL MoS$_2$) as the BGC and further assume free-standing ML and BL MoS$_2$ share identical BGC. The calculated work function of free-standing BL MoS$_2$ at the DFT level is 5.04 eV, which is 0.21 eV slightly smaller than its experimental value. The calculated work function of free-standing ML MoS$_2$ at the DFT level is merely 0.08 eV greater than the calculated one of free-standing BL MoS$_2$.

### D. Schottky barrier at ML and BL MoS$_2$-metal interfaces

The band structure of ML MoS$_2$ is moderately destroyed by Ag, Pt, and Ni surfaces, and the vertical *n*-type (or *p*-type) Schottky barrier for this medium bonding case can be obtained from the difference between $E_f$ of the interfacial system and the CBM (or VBM) of ML MoS$_2$. By contrast, a strong band hybridization has taken place for 2D MoS$_2$ on Sc and Ti surfaces, resulting in metallization of 2D MoS$_2$ and absence of vertical Schottky barrier for the five contacts. It has been proved that for the semiconductor fully under metal (namely, in the electrode region), the many-electron effects are greatly depressed if a charge transfer takes place [38,53]; as a result, the KS band edge positions and band gap are a good approximation. Therefore, we only consider many-electron effects for the band structure of the semiconductor in the channel of a device in the case that a metallization takes place between metal and underneath 2D MoS$_2$. Namely, only as calculating the lateral SBH of 2D MoS$_2$-Sc and -Ti contacts, we consider many-electron effects. We have electron SBH $\Phi_V^{DFT}$ = 0.212 and 0.633 eV for Ag and Ni contacts, respectively, and hole SBH $\Phi_V^{DFT}$ = 0.520 eV for Pt contact. While for Au contact, the band structure of ML MoS$_2$ is identifiable, and $E_f$ is nearly in the middle of the band gap. Therefore, ML MoS$_2$-Au contact has a midgap Schottky barrier, and this is consistent with the experiment [54].

Lateral Schottky barrier $\Phi_L$ is determined by the energy difference between $E_f$ of the interfacial system and the CBM (*n*-type) or VBM (*p*-type) of channel ML MoS$_2$. ML MoS$_2$ forms an Ohmic contact with Sc in the lateral direction at the DFT level since $E_f$ of the interfacial system is higher than the $E_{CBM}^{DFT}$ of channel MoS$_2$. However, there is a large lateral SBH at the *GW* level, with $\Phi_L^{GW}$ = 0.539 eV. There is a lateral *n*-type Schottky barrier for Ti contacts at both the DFT and *GW* levels, with smaller $\Phi_L^{DFT}$ = 0.216 and larger $\Phi_L^{GW}$ = 0.796 eV. The DFT SBHs of ML MoS$_2$-metal interfaces are in good agreement with the previous DFT calculations (see Table 1). For example, the lateral DFT SBH for ML MoS$_2$-Ti interface is 0.33 eV calculated by Banerjee *et al*. [23,39]. There is some uncertainty in identifying the metallization. However, even if we identify a metallization for ML MoS$_2$ under Ag, Pt, and Ni, the values of the resulting lateral SBHs are close to those of the vertical SBHs.

The vertical Ohmic contact feature remains on Sc and Ti surfaces from ML to BL MoS$_2$, because the strong band hybridization remains. From ML to BL MoS$_2$, $\Phi_V$ in MoS$_2$-Au contact is significantly decreased by 0.096 eV at the DFT level as a result of the reduction of the band gap (0.220 eV). The vertical SBHs for Ag, Pt, and Ni contacts are slightly decreased by 0.074, 0.175, and 0.021 eV, respectively, at the DFT level from ML to BL MoS$_2$. The reduced SBH from ML to BL MoS$_2$ is in good agreement with the experiment [21]. BL MoS$_2$ still forms an Ohmic contact with Sc in the lateral direction.

Since the lattice mismatches are large for the Sc-MoS$_2$ (4.485%) and Ti-MoS$_2$ (6.791%) interfaces in the above study, we further enlarge the supercell to reduce the lattice mismatch. The $\sqrt{13}\times\sqrt{13}$ unit cell of MoS$_2$ is adjusted to the $2\sqrt{3}\times 2\sqrt{3}$ unit cells of Sc(0001) surface, with the lattice mismatch decreased to 1.29%. The $2\sqrt{3}\times 2\sqrt{3}$ unit cell of MoS$_2$ is adjusted to $\sqrt{13}\times\sqrt{13}$ unit cells of Ti(0001) surface, with the lattice mismatch decreased to 2.99%. Compared with the large mismatch configuration, the small mismatch ones do not change the contact type and just slightly increase $\Phi_L^{DFT}$ from 0.187 (0.096) to 0.216 (0.161) eV for ML (BL) MoS$_2$-Ti contact, which is closer to a DFT value of 0.33 eV of Banerjee *et al*. [23,39] based on a larger ML MoS$_2$-Ti interfacial supercell containing 6 Mo and 12 S atoms per unit cell in the contact region.

The experimentally extracted SBHs of ML and BL MoS$_2$-Ti contact are 0.3~0.35 [23]

and 0.065 eV [26], respectively, which are in agreement with our calculated values of 0.216 and 0.161 eV at the DFT level but apparently deviate from the corresponding values with many-body effect correction (0.796 and 0.341 eV, respectively). Such a comparison suggests that many-electron effects have been greatly depressed by the charge transfer between channel $MoS_2$ and the electrodes, which significantly screens the electron-electron Coulomb interaction and validates sing-electron approximation. In other word, the transport gap of ML $MoS_2$ is determined by the DFT band gap rather than the quasiparticle band gap.

In a recent work, the SBH and the transport gap of phosphorene have been measured [55]. Phosphorene is *p*-type doped by Ni electrodes, and the transport gaps of ML and few layer phosphorene with Ni electrodes are in good agreement with the DFT band gaps at the GGA level. For example, the transport gap of ML and BL phosphorene are 0.98 ± 0.4 and 0.71 ± 0.4 eV [55], respectively, and the corresponding band gaps are 0.91 and 0.6 eV at the DFT level [56], while the quasiparticle band gaps are 2.0 and 1.3 eV [57]. The measured *p*-type SBH of ML phosphorene is 0.34 ± 0.2 eV, which is also in good agreement the one at the DFT level (0.26 eV) [58] but much smaller than the one at the *GW* level (0.82 eV) [59]. Therefore, the suppressed many-electron effects can be expanded to a general device if 2D channel semiconductor is doped by electrodes, and correspondingly the transport gap depends on the DFT band gap instead of the quasiparticle band gap.

In our above calculations, we adapt the lattice constant of $MoS_2$ to that of metal surfaces as the match way in Ref. [27] in view of the fact that the bulk metal electrode is more robust than ML and BL $MoS_2$. We note that the lattice constant of $MoS_2$ is fixed in Ref. [28]. In order to explore the effects of the match way on the work function of $MoS_2$-metal interface, we give the work function of interfacial systems in the case of fixing $MoS_2$ lattices in Table 2. The work function of the system with Ti surface adjusted to $MoS_2$ is 0.205 eV smaller than that of the system with BL $MoS_2$ adjusted to Ti surface; consequently, the lateral SBH disappears. Such a result is in consistent with the experimental SBH of 0.065 eV for BL $MoS_2$-Ti contact [26]. There are nearly no difference in work function between these two strained method for BL $MoS_2$-Sc and ML $MoS_2$-Ti contacts.

### E. Tunneling barrier at ML and BL $MoS_2$-metal interfaces

In order to complete the analysis of contacts, we further investigate the electrostatic potential at the ML MoS$_2$-metal interfaces and show the results in Fig. 6. The tunneling barrier $\Delta V$ here is defined as the potential energy above the Fermi energy between the MoS$_2$ and metal surfaces, indicated by the black rectangular, and the tunneling width $w_B$ is defined as the full width at half maximum of the $\Delta V$. As shown in Fig. 6 and Table 3, the $\Delta V$ values at the strong hybridization interfaces (Sc, Ti, Ni, and Pt) are significantly lower and the $w_B$ values are significantly narrower than those at the weak ones (Ag and Au). A lower barrier height and a narrower width at a semiconductor-metal interface mean a higher electron injection efficiency. We estimate the tunneling probabilities $T_B$ from metal to MoS$_2$ using a square potential barrier model as:

$$T_B = \exp(-2 \times \frac{\sqrt{2m\Delta V}}{\hbar} \times \omega_B) \tag{3}$$

where $m$ is the effective mass of a free electron and $\hbar$ is the Plank's constant. The $T_B$ values are thus estimated to be 100, 100, 74.33, 53.21, 19.68, and 4.74% for Sc, Ti, Ni, Pt, Ag, and Au contacts, respectively (see Table 3). Apparently, Sc and Ti contacts have perfect transmission. The tunneling properties of the tunneling barrier at the BL MoS$_2$-metal interfaces are also summarized in Table 3. Compared with the case of ML MoS$_2$ contact metals, there is little change in the tunneling properties for BL MoS$_2$, indicating that the tunneling properties are insensitive to the MoS$_2$ layer number.

In the light of Schottky barrier and tunneling barrier, the nature of MoS$_2$- metal contact can be classified into five types. Sc can form high quality contact interface with ML and BL MoS$_2$ with zero tunneling barrier and zero Schottky barrier, leading to Ohmic contact (Figs. 7(d) and 7(i)). Although the metallization of ML MoS$_2$ with Ti eliminates the Schottky barrier at the interface B, the injected electrons from the metal still confront a *n*-type Schottky barrier at the interface D, leading to Type 1 in Fig. 7(e). The nature of BL MoS$_2$-Ti contacts also belong to Type 1. It is worthwhile to note that the tunneling barrier vanishes in Type 1 contact due to the metallization at interface B. Unlike the case in Type 1, there is a tunneling barrier at the interface B in Types 2 and 3 contacts. Only *p*-type Schottky barrier forms in ML and BL MoS$_2$-Pt contacts (Type 3, Figs. 7(g) and 7(l)). In Type 4 contact (ML and BL MoS$_2$-Au), Schottky barrier and tunneling barrier are formed at the interface B, and SBH is zero at the interface D because of the lack of orbital overlaps.

### F. Fermi level line-up

Our calculated $\Phi_{SB}$ of ML MoS$_2$ on all investigated metal surfaces are listed in Fig. 8, in which the *GW* results for Sc and Ti contacts are also provided for comparison. For Sc contacts, the $\Phi_{SB}$ obtained by transport calculations is also presented, which will be discussed later in the transport properties. The CBM of Sc, Ti, Ag, Au, and Ni-ML MoS$_2$ systems are closer to the Fermi levels than the VBM, leading to the *n*-type contact. While the VBM of Pt-ML MoS$_2$ absorbed system is closer to the Fermi levels and form *p*-type contact. The *n*-type characteristic of ML MoS$_2$ on Sc, Ti, Ag, Au, and Ni surfaces has been observed experimentally [8,21,24], and the *p*-type characteristic on Pt surface is calculated in other DFT calculation [28]. Therefore, ML MoS$_2$ *p–n* junction can be fabricated by using Sc, Ti, Ag, Au, or Ni to contact one end of ML MoS$_2$ and Pt to contact the other end of it. The ML MoS$_2$ *p–n* junction can be used to develop optoelectronics or valley-optoelectronics technology [60]. Comparing the $\Phi_{SB}$ at the DFT and *GW* levels for Sc and Ti contacts, we find that the doping type is unchanged.

The calculated $\Phi_{SB}$ of BL MoS$_2$ on the six metal surfaces are listed in Fig. 9, in which the *GW* results for Sc and Ti contacts are also provided for comparison. Compared with Fig. 8, the *GW* correction to the band gap of BL MoS$_2$ is less significant because many-body effect is reduced with the increasing size in the vertical direction. BL MoS$_2$ FET is also *p*-type doped by Pt contact and *n*-type doped by the other five contacts. Consistently, the experiments have found *n*-type characteristic of BL MoS$_2$ on Ti and Au surfaces [11,26].

The Fermi level shift $\Delta E_f$ is defined as the difference between the interfacial systems and free-standing 2D MoS$_2$ work functions when the band hybridization occurs at the interfaces (Sc, Ti, Ni, Pt, and Ag contacts). $\Delta E_f$ is defined as $\Delta E_f = E_{mid} - E_f$ for the interface without band hybridization (for Au contact), where $E_{mid}$ is the midpoint of the identifiable band gap of MoS$_2$. Negative (positive) $\Delta E_f$ means *n*-type (*p*-type) doping of 2D MoS$_2$. The doping types determined from $\Delta E_f$ are in agreement with those determined from Figs. 8 and 9. $\Delta E_f$ as a function of the difference between the clean metal ($W_M$) and ML (BL) MoS$_2$ work functions ($W_{MoS_2}$) is shown in Fig. 10 (11). The cross point from *n*- to *p*-type doping is $W_M - W_{MoS_2} =$ 0.21 (0.13) eV for ML (BL) case. The $\Delta E_f$ has a nearly linear dependence with the $W_M -$ $W_{MoS_2}$ with a slope of 0.64 in both ML and BL MoS$_2$-metal contacts, approximately equal to the previously reported theoretical value of 0.71 in ML MoS$_2$-metal contacts [28]. Notably, $\Delta E_f$ is insensitive to the MoS$_2$ layer number, leading to the same linear relation between $\Delta E_f$ and work function. Note that the slope close to 0 indicates a strong Fermi level pinning, and we therefore observe a partial Fermi level pinning picture once more when the six metals contact ML MoS$_2$. The partial Fermi level pinning is a synergic result of the metal work function modification and the interface gap states formation in the studied interface systems [28].

### G. Quantum Transport Simulation

We note that the experiment reported that few layer (3-18 layers) MoS$_2$-Sc contact still has a very small SBH of 0.03 eV [21]. ML and BL MoS$_2$ should have a larger SBH due to the enhanced band gaps compared with few layer MoS$_2$ and this is inconsistent with the predicted Ohmic contact for Sc electrode ($E_f$ of Sc electrode is above the CBM of channel ML/BL MoS$_2$ by 0.22/0.21 eV) in dual interface model calculation. In the dual interface model, one determines the SBH indirectly by calculating the work functions of MoS$_2$ under metal and channel MoS$_2$ separately. In a real device, there is possible complex coupling between MoS$_2$ under metal and channel MoS$_2$ (namely Fermi level pinning). A direct and better theoretical approach to determine the SBH of a FET is to calculate the transport property of a FET of a two-probe model by using the DFT method coupled with NEGF.

In our quantum transport simulations, the device is constructed of ~ 60 Å ML/BL MoS$_2$ in the channel region along the transport direction and ML/BL MoS$_2$-Sc interfaces in the electrode region. The lattice constant of the ML/BL MoS$_2$ should be carefully chosen, as it directly affects the size of the band gap and thus transport properties. In a real device, the lattice constant of the ML/BL MoS$_2$ in the central region is close to that of free-standing ML/BL MoS$_2$, while in the electrode region the lattice constant of the ML/BL MoS$_2$ should be adapted to that of corresponding bulk metals supercell. In order to capture this feature, we consider two extreme cases in the transport calculations: in Model I, the lattice constant of ML/BL MoS$_2$ in the device is adapted to that of Sc surface, and in Model II, the lattice constant of Sc surface is adapted to that of ML/BL MoS$_2$. One could expect that the transport properties of the real device should be between the two cases.

The transmission spectra of ML and BL MoS$_2$ transistors using the two models calculated with SZP basis set are provided in Fig. 12(a), respectively. A test shows that a larger DZP basis set gives a quite close SBH. The transmission spectra of ML MoS$_2$ transistors give transport gaps of 0.92 eV in Model I and 1.67 eV in Model II, and the latter value is quite close to the band gap (1.68 eV) of free-standing ML MoS$_2$. The Fermi level $E_f$ is slightly below the CBM in both two models, showing a *n*-type Schottky barrier between ML MoS$_2$ and Sc electrode in the devices. The values of the electron SBH are read as 0.040 eV and 0.260 eV in Models I and II, respectively. As the real system is between the two extreme cases, we estimate the SBH in the real ML MoS$_2$ transistor with Sc electrodes to be around 0.150 eV by roughly averaging the values of the two cases. As the number of MoS$_2$ layers increases, its band gap decreases. Our transport simulations also show a reduction (~ 0.09 eV in Model I and 0.56 eV in Model II, respectively) of the transport gap of BL MoS$_2$ compared to that of ML. The average value of SBH in the BL MoS$_2$ with Sc contact over the two models is estimated to 0.185 eV in the transport simulation.

The local density of states (LDOS) versus the coordinate along the transport direction of ML MoS$_2$ tansistors using the two models calculated with SZP basis set are provided in Figs. 12(b-c). It is apparent from the LDOS that the conduction band is bent downward due to an electron transfer from Sc to channel ML MoS$_2$ where no impurity states exist. Such a downward bending is different from a common band upward bending in a metal-*n* type semiconductor interface where donor states exist and electrons are transferred from semiconductor to metal. In accordance with the value calculated from the transmission spectra, the LDOS also shows an average *n*-type SBH of 0.15 eV for ML MoS$_2$-Sc interface. Taken together, unlike the DFT energy band analysis, which gives a Ohmic contact, the quantum transport simulations give a *n*-type Schottkey contact for ML and BL MoS$_2$ Sc-interfaces with electron SBH of 0.150 and 0.185 eV, respectively, which are in agreement with the experiment [21], where 3-18 layer MoS$_2$ Sc-interface has an electron SBH of 0.03 eV.

The failure of the energy band analysis in predicting MoS$_2$-Sc contact comes from the ignorance of the coupling between MoS$_2$ under Sc and channel MoS$_2$ because we calculate the electronic properties of the electrode and the channel region separately during deriving the lateral SBH. This coupling makes the Ohmic contact difficult to occur because the Fermi level is pinned below the CBM of $MoS_2$. Therefore, caution must be taken for any lateral Ohmic contact predicted by the energy band analysis, and a further quantum transport calculation is necessary to obtain a more reliable interface. Actually, the Ohmic contact in ML phosphorene-Pd contact derived from the energy band analysis also turns out to be artificial in terms of the quantum transport simulations [58].

If the SBH appears in the vertical direction, the coupling between metal and $MoS_2$ has been taken into account in the energy band calculations because the metal and semiconductor parts are treated a whole. In this case, it appears that the quantum transport simulation will give similar SBH. We calculate the transport properties of ML and BL $MoS_2$ with Pt electrodes (the channel length is ~ 63 Å). As the lattice mismatch between $MoS_2$ and Pt supercell is small (~1.2 %), we only consider Model I in which the lattice constant of ML/BL $MoS_2$ is adapted to that of Pt supercell. As shown in Fig. S1, transport gaps of 1.34 and 1.03 eV are observed for ML and BL $MoS_2$-Pt interfaces in the transmission spectra. The Fermi level in BL $MoS_2$-Pt contact is apparently closer to the VBM, having a hole SBH of 0.32 eV, which is in indeed good agreement with the one (0.345 eV) from the energy band analysis. It appears that the coupling between Pt and BL $MoS_2$ has been taken into account in the energy band calculations. It is notable that energy band analysis and the quantum transport simulations also give similar *p*-type SBHs for Pt-$WSe_2$ interface (0.28 and 0.34 eV, respectively) [61]. However, the Fermi level of ML $MoS_2$-Pt contact is located nearly in the middle of the transport gap (slightly closer to the CBM of ML $MoS_2$), showing a midgap SBH. This is not in consistent with the energy band analysis, which favors a *p* doping of ML $MoS_2$ (the hole SBH is 0.520 eV). The story becomes more complicated as the experimental observations show electron SBH of ~ 0.23 eV for 3-8 layers $MoS_2$-Pt interface [21]. The origin of the controversy among the energy band analysis, quantum transport simulations, and experiments remains unclear and more studies on the $MoS_2$-Pt system are desirable. It is interesting to mention that, in the $MoS_2$-Pd system, both *n*- and *p*-doped characteristics of $MoS_2$ have been reported [62,63,64]. It is well known that Pt has a larger work function than Pd (6.1 eV vs 5.6 eV) [48], and generally Pt can induce *p* doping more easily. It appears that the possibility of *p* doping of ML and BL $MoS_2$ by Pt contact cannot be excluded completely.

### H. Discussions

There are four types of commonly used band gap for a 2D semiconductor: transport gap, quasiparticle gap, optical gap (dominated by strong exciton effects), and DFT gap. Taking ML/BL phosphorene as an example, the four band gaps are: 0.98/0.71 [55], 2.0/1.3 [56], 1.30/0.70 , and 0.91/0.60 eV [57]. Apparently, the DFT band gap and optical gap are the closest to the transport gap because the 2D channel semiconductor is doped by carrier. In addition to doping by electrode, the 2D semiconductor channel is also subject to electrostatic doping by gate. This is another cause why many-electron effects are strongly depressed of the 2D channel semiconductor. However, the transport gaps are still about 10% slightly larger than their respective DFT gaps in phosphorene [59], suggestive of the existence of weak many-electron effects with about 10% correction in doped phosphorene, which is one order of magnitude smaller than that in intrinsic phosphorene. Actually, the band gap of a heavily doped silicene is 0.34 and 0.38 eV at the DFT and *GW* level, respectively, consistent with a correction of about 10% upon the inclusion of the many-body effects [38]. From a physical point of view, the transport gap of a 2D semiconductor should equal to the quasi-partical band gap of heavily doped system, which is slightly larger than the DFT band gap. Fig. 13 illustrates the size relation of these common band gaps. Hence, a small correction (increase by about 10%) to the DFT CBM and VBM appears required to obtain the accurate CBM and VBM positions of a doped 2D semiconductor and thus the accurate SBH at the interface.

### IV. CONCLUSION

In summary, we provide the first comparative study of the interfacial properties of monolayer and bilayer $MoS_2$ on Sc, Ti, Ag, Pt, Ni, and Au surfaces by using different theoretical approaches. A comparison between the calculated and observed Schottky barrier heights suggests that many-electron effects are strongly depressed (but not vanished) and the transport gap of a device depends on the DFT band gap (a minor correction is still needed) rather than the quasiparticle band gap. Such a depression of many-electron effects can be applied to a general metal-2D semiconductor interface. Schottky barrier heights are decreased from ML $MoS_2$-metal interfaces to BL $MoS_2$-metal interfaces due to the interlay coupling, implying that BL $MoS_2$ with a higher electron injection efficiency is more suitable for a transistor than ML MoS$_2$. Most strikingly, we find that DFT energy band calculations are unable to reproduce the experimental Schottky barrier heights in some cases and give incorrect Ohmic contact prediction because the Fermi level pinning has not been fully taken into account. In the interface study between other 2D material and metal, such a shortcoming remains. To solve such a problem, a higher level ab initio quantum transport calculation based on a two-probe model is desired.

**ACKNOWLEDGMENT**

This work was supported by the National Natural Science Foundation of China (No. 11274016/11474012), the National Basic Research Program of China (No. 2013CB932604/ 2012CB619304), and the National Science Foundation Grant (1207141).

**Author contribution**

The idea was conceived by J. L. The DFT electronic band calculation was performed by H. Z, and the device simulation was performed by R. Q and M. Y.. The data analyses were performed by J. L., H. Z., R. Q., J. S., L. Y., J. Y., Z. N., Y. W., M. Y., Z. S., and Y. P. This manuscript was written by H. Z, R. Q., and J. L. All authors contributed to the preparation of this manuscript.

**Table 1.** Calculated interfacial properties of ML and BL MoS$_2$ on metal surfaces. The experimental cell parameters of the surface unit cells shown in Fig. 1 for various metals are given under the metals. The corresponding lattice mismatches are given. The equilibrium distance $d_{S\text{-}M}$ is the averaged distance between the surface S atoms of MoS$_2$ and the relaxed positions of the topmost metal layer in the $z$ direction. $E_b$ is the binding energy per surface S atom between MoS$_2$ and a given surface. $W_M$ and $W$ are the calculated work functions for clean metal surface and metal surface adsorbed by MoS$_2$, respectively. $\Phi_v$ and $\Phi_L$ are the vertical and lateral SBH at the DFT level, respectively, of a MoS$_2$ transistor (see Fig. 7(c)); the SBH obtained in other DFT calculations, the $GW$-corrected SBHs, and the measured SBH are given below them in parenthesis for comparison. $\Delta E_f$ is the Fermi level shift of 2D MoS$_2$. The corresponding values for Sc and Ti surfaces in small mismatch are also given. Caution must be taken for the data of ML and BL MoS$_2$-Ni contacts due to the large lattice mistmach (9.1%) limited by the computational resource.

| Metal | Mismatch | $W_M$ (eV) | ML MoS$_2$ | | | | | | BL MoS$_2$ | | | | | |
|---|---|---|---|---|---|---|---|---|---|---|---|---|---|---|
| | | | $d_{S\text{-}M}$ (Å) | $E_b$ (eV) | $W$ (eV) | $\Delta E_f$ (eV) | $\Phi_v$ (eV) | $\Phi_L$ (eV) | $d_{S\text{-}M}$ (Å) | $E_b$ (eV) | $W$ (eV) | $\Delta E_f$ (eV) | $\Phi_v$ (eV) | $\Phi_L$ (eV) |
| Sc 3.308 (Å) | 4.485% | 3.593 | 1.786 | 1.181 | 4.369 | -0.881 | 0.000 | 0.000 | 1.783 | 1.182 | 4.306 | -0.944 | 0.000 | 0.000 |
| | | | | | | | 0.000 | (0.539)$^{GW}$ | | | | | 0.000 | (0.000)$^{GW}$ |
| | 1.290% | | | | 4.192 | | | 0.000 | | | 4.300 | -0.950 | | 0.000 |
| | | | | | | | | (0.362)$^{GW}$ | | | | | | (0.000)$^{GW}$ |
| Ti 2.951 (Å) | 6.791% | 4.427 | 1.557 | 1.812 | 4.597 | -0.653 | 0.000 | 0.187 | 1.560 | 1.848 | 4.616 | -0.634 | 0.000 | 0.096 |
| | | | | | | | — | (0.3-0.35)$^b$ | | | | | — | (0.065)$^d$ |
| | | | | | | | (0.000)$^a$ | (0.33)$^c$ | | | | | — | (0.276)$^{GW}$ |
| | | | | | | | — | (0.731)$^{GW}$ | | | | | 0.000 | — |
| | 2.990% | | | | 4.626 | | 0.000 | 0.216 | | | 4.681 | | | 0.161 |
| | | | | | | | | (0.796)$^{GW}$ | | | | | | (0.341)$^{GW}$ |
| Ag 5.778(Å) | 5.367% | 4.489 | 2.961 | 0.503 | 4.662 | -0.588 | 0.212 | 0.000 | 2.917 | 0.547 | 4.763 | -0.487 | 0.138 | 0.000 |
| Ni 4.984(Å) | 9.112% | 5.222 | 2.094 | 0.830 | 5.001 | -0.249 | 0.633 | 0.000 | 2.097 | 0.729 | 5.102 | -0.148 | 0.612 | 0.000 |
| Au 5.768(Å) | 5.185% | 5.226 | 3.405 | 0.307 | 5.173 | -0.077 | 0.763 | 0.000 | 3.325 | 0.354 | 5.187 | -0.063 | 0.667 | 0.000 |
| | | | | | | | (0.88)$^e$ | (0.000)$^e$ | | | | | — | — |
| Pt 5.549(Å) | 1.191% | 5.755 | 2.476 | 0.570 | 5.444 | 0.194 | 0.520 | 0.000 | 2.438 | 0.634 | 5.476 | 0.226 | 0.345 | 0.000 |
| | | | | | | | (0.770)$^f$ | | | | | | | |

$^a$ DFT values from Refs. [22,39].
$^b$ Experimental value [23].
$^c$ DFT value from Ref. [23,39].
$^d$ Experimental value at a low temperature in Ref. [26].
$^e$ DFT value from Ref. [28].
$^f$ The SBH for ML MoS$_2$-Pt is hole SBH and the DFT value from Ref. [28].

**Table 2.** The work functions $W$ for metal surface adsorbed by MoS$_2$. The vertical $\Phi_v$ and lateral $\Phi_L$ SBH at the DFT level of a MoS$_2$ transistor (see Fig. 7(c)) when the lattice constants of metal surfaces are adjusted to that of MoS$_2$.

| Metal | ML MoS$_2$ | | | BL MoS$_2$ | | |
|---|---|---|---|---|---|---|
| | $W$ (eV) | $\Phi_v$ (eV) | $\Phi_L$ (eV) | $W$ (eV) | $\Phi_v$ (eV) | $\Phi_L$ (eV) |
| Sc | 4.161 | 0.000 | 0.000 | 4.310 | 0.000 | 0.000 |
| Ti | 4.561 | 0.000 | 0.151 | 4.411 | 0.000 | 0.000 |
| Ag | 4.625 | 0.215 | 0.000 | 4.558 | 0.138 | 0.000 |
| Ni | 5.115 | 0.705 | 0.000 | 5.240 | 0.820 | 0.000 |
| Au | 5.058 | 0.813 | 0.000 | 5.211 | 0.559 | 0.000 |
| Pt | 5.432 | 0.648 | 0.000 | 5.079 | 0.659 | 0.000 |

**Table 3.** Tunneling barrier height $\Delta V$, width $w_B$, and probabilities ($T_B$) through the ML (BL) MoS$_2$-metal interfaces.

| Metal | ML MoS$_2$ | | | BL MoS$_2$ | | |
|---|---|---|---|---|---|---|
| | $\Delta V$ (eV) | $w_B$ (Å) | $T_B$ (%) | $\Delta V$ (eV) | $w_B$ (Å) | $T_B$ (%) |
| Sc | 0.000 | 0.000 | 100 | 0.000 | 0.000 | 100 |
| Ti | 0.000 | 0.000 | 100 | 0.000 | 0.000 | 100 |
| Ag | 3.003 | 0.916 | 19.68 | 2.911 | 0.904 | 20.61 |
| Ni | 0.785 | 0.327 | 74.33 | 0.822 | 0.336 | 73.20 |
| Au | 4.697 | 1.374 | 4.74 | 4.585 | 1.356 | 5.11 |
| Pt | 1.810 | 0.458 | 53.21 | 1.871 | 0.517 | 48.47 |

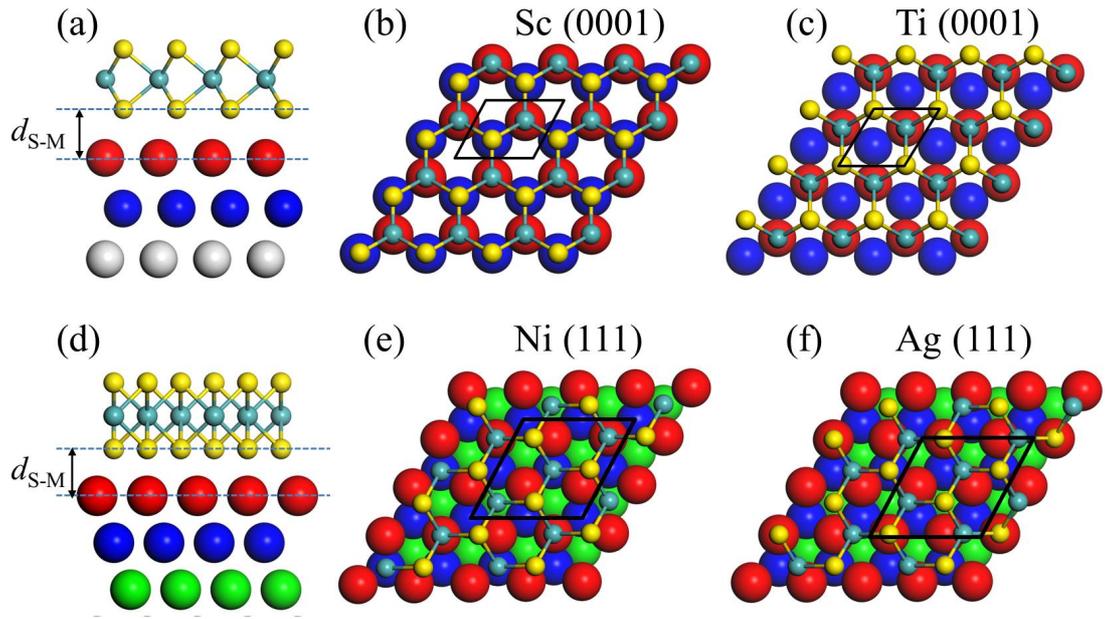

**FIG. 1.** Interfacial structures of the most stable configuration for ML MoS$_2$ on metal surfaces. (a) Side and (b) top views of ML MoS$_2$ on Sc(0001) surface. (c) Top view of MoS$_2$ on Ti(0001) surface. (d) Side and (e) top views of ML MoS$_2$ on Ni and Pt(111) surfaces. (f) Top view of MoS$_2$ on Ag and Au(111) surfaces. $d_{S-M}$ is the equilibrium distance between the metal surface and the bottom layer MoS$_2$. The rhombi plotted in black line shows the unit cell for each structure.

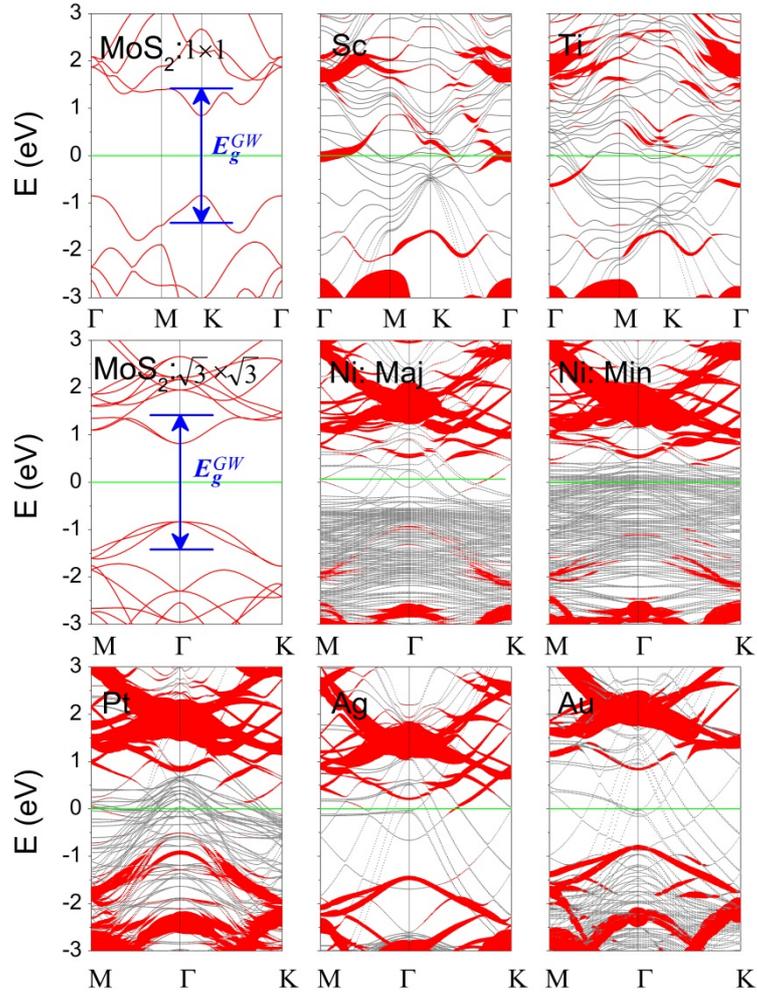

**FIG. 2**. Band structures of ML MoS$_2$ on Sc, Ti, Ni, Pt, Ag, and Au surfaces by the DFT method, respectively. The Fermi level is at zero energy. Gray line: metal surface bands; red line: bands of MoS$_2$. The line width is proportional to the weight. Blue line: the positions of CBM and VBM of MoS$_2$ after the *GW*-BGC correction. The labels Maj/Min indicate the majority-spin and minority-spin bands of MoS$_2$ on Ni surface. The band structure of free-standing ML MoS$_2$ calculated in a primitive unit cell and a $\sqrt{3}\times\sqrt{3}$ supercell are provided for comparison.

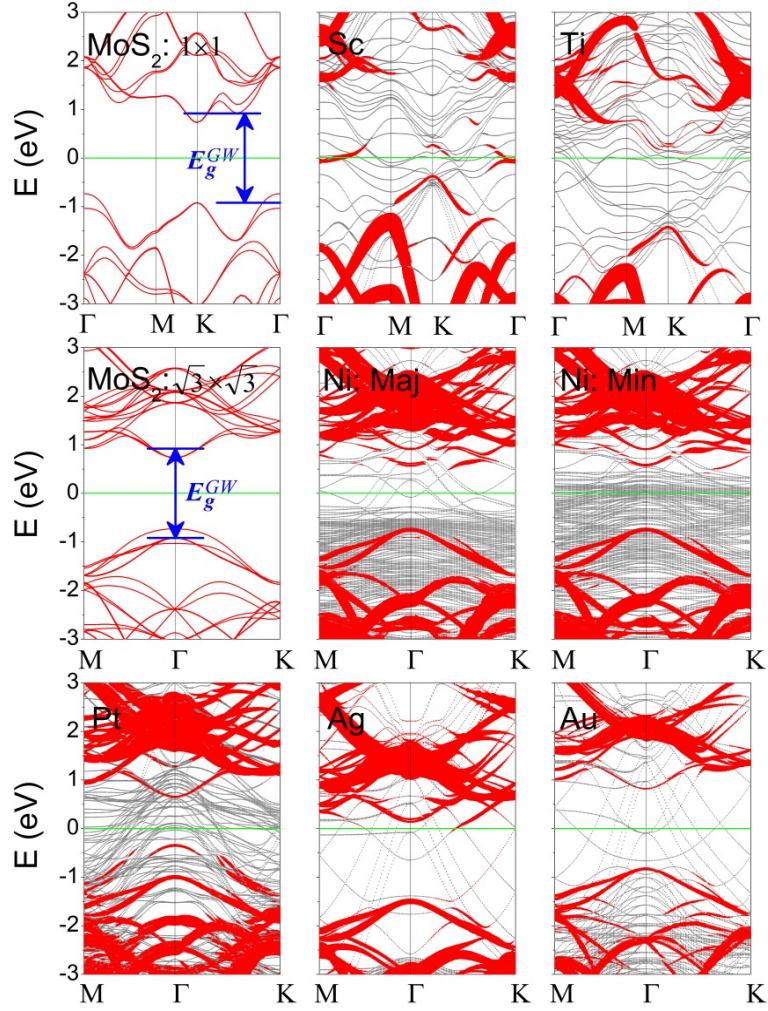

**FIG. 3**. Band structures of BL MoS$_2$ on Sc, Ti, Ni, Pt, Ag, and Au surfaces by the DFT method, respectively. The Fermi level is at zero energy. The gray (red) line denotes metal surface (BL MoS$_2$) bands. The line width is proportional to the weight. Blue line: the positions of CBM and VBM of MoS$_2$ at the *GW*-BGC scheme. The labels Maj/Min indicate the majority-spin and minority-spin bands of BL MoS$_2$ on Ni surface. The band structure of free-standing BL MoS$_2$ calculated in a primitive unit cell and a $\sqrt{3}\times\sqrt{3}$ supercell are provided for comparison.

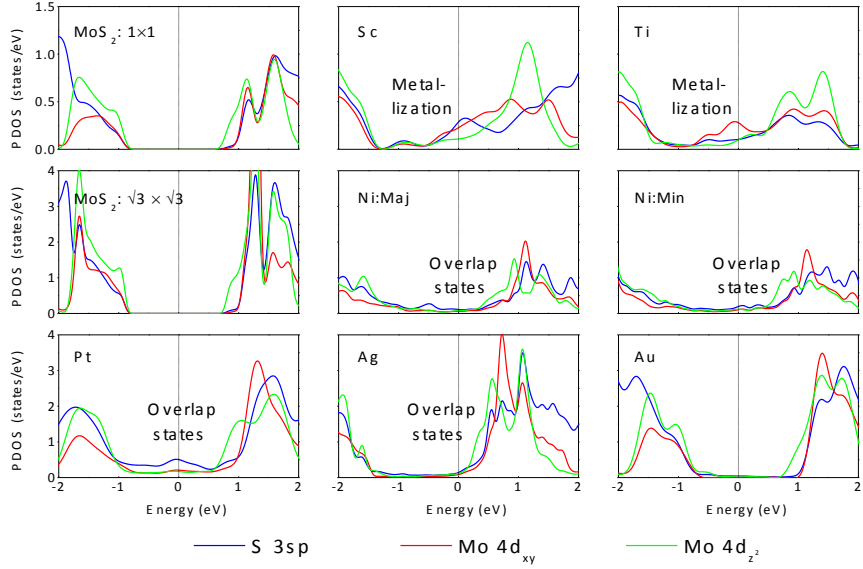

**FIG. 4.** Partial density of states (PDOS) (DOS on specified atoms and orbitals, for example, Mo-*d* (*d*-orbital on Mo)) of ML $MoS_2$ on Sc, Ti, Ni, Pt, Ag, and Au surfaces at the DFT level. The Fermi level is at zero energy. The PDOS of free-standing ML $MoS_2$ calculated in a primitive unit cell and a $\sqrt{3}\times\sqrt{3}$ supercell is provided for comparison.

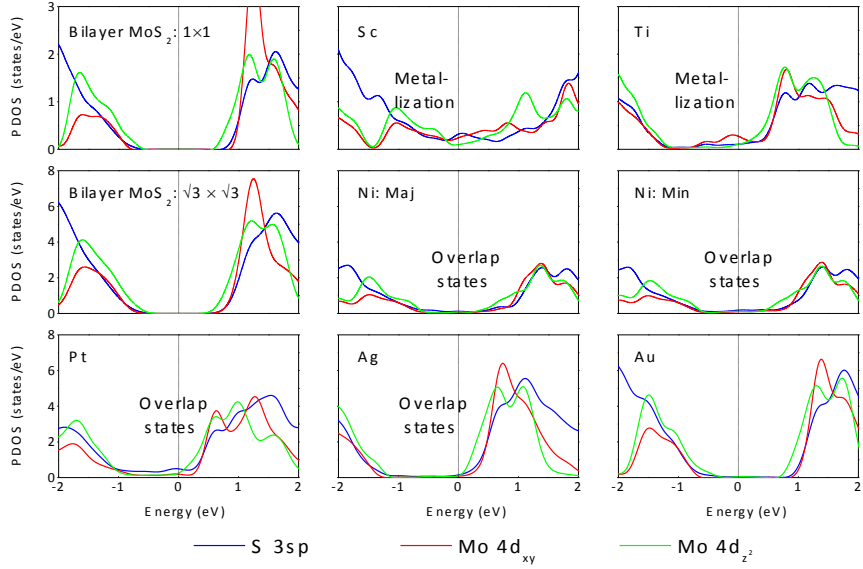

**FIG. 5.** PDOS of BL MoS$_2$ on Sc, Ti, Ni, Pt, Ag, and Au surfaces at the DFT level. The Fermi level is at zero energy. The PDOS of free-standing BL MoS$_2$ calculated in a primitive unit cell and a $\sqrt{3}\times\sqrt{3}$ supercell are provided for comparison.

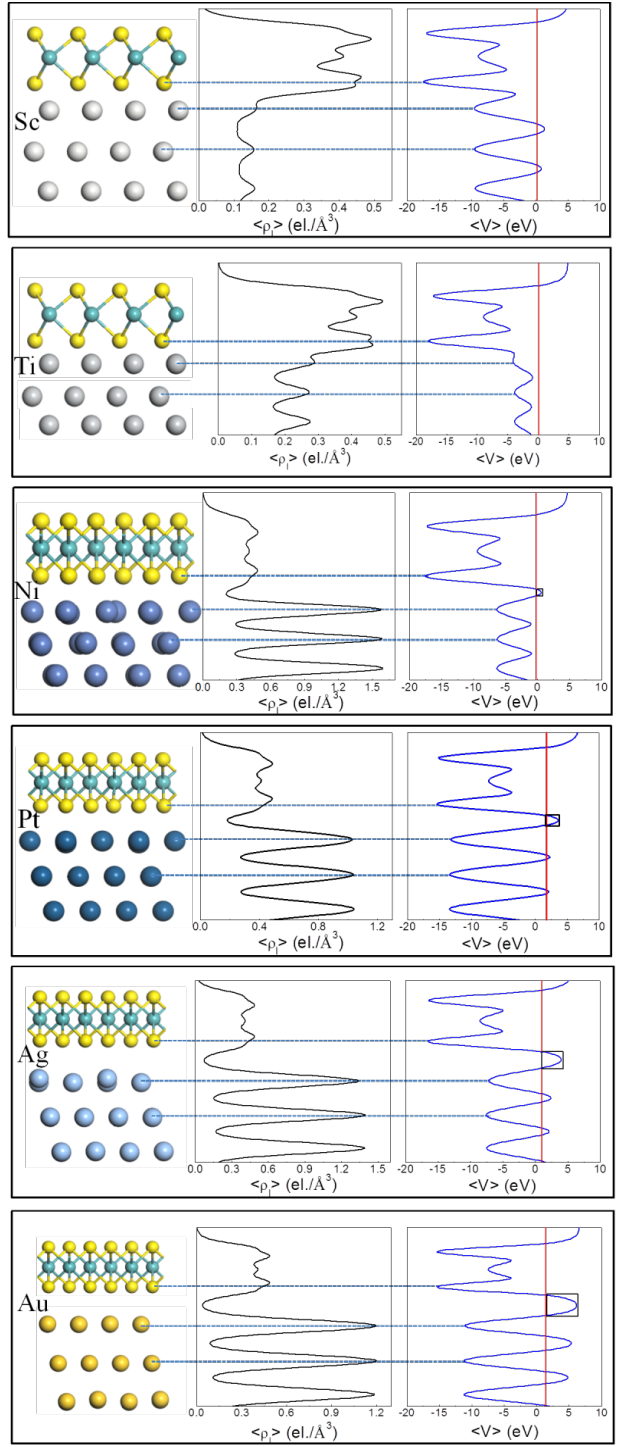

**FIG. 6.** Electronic structure at the interface between ML $MoS_2$ and metal at the DFT level. $<\rho_l>$ is the average value in planes parallel to the interface of $MoS_2$-metal. $<V>$ is the average electrostatic potential in planes normal to the $MoS_2$-metal interface. The dot lines indicate the location of the sulfur layer and the metal layers at the interface. The higher the $\rho_l$ at the interface is, the higher the electron injection is.

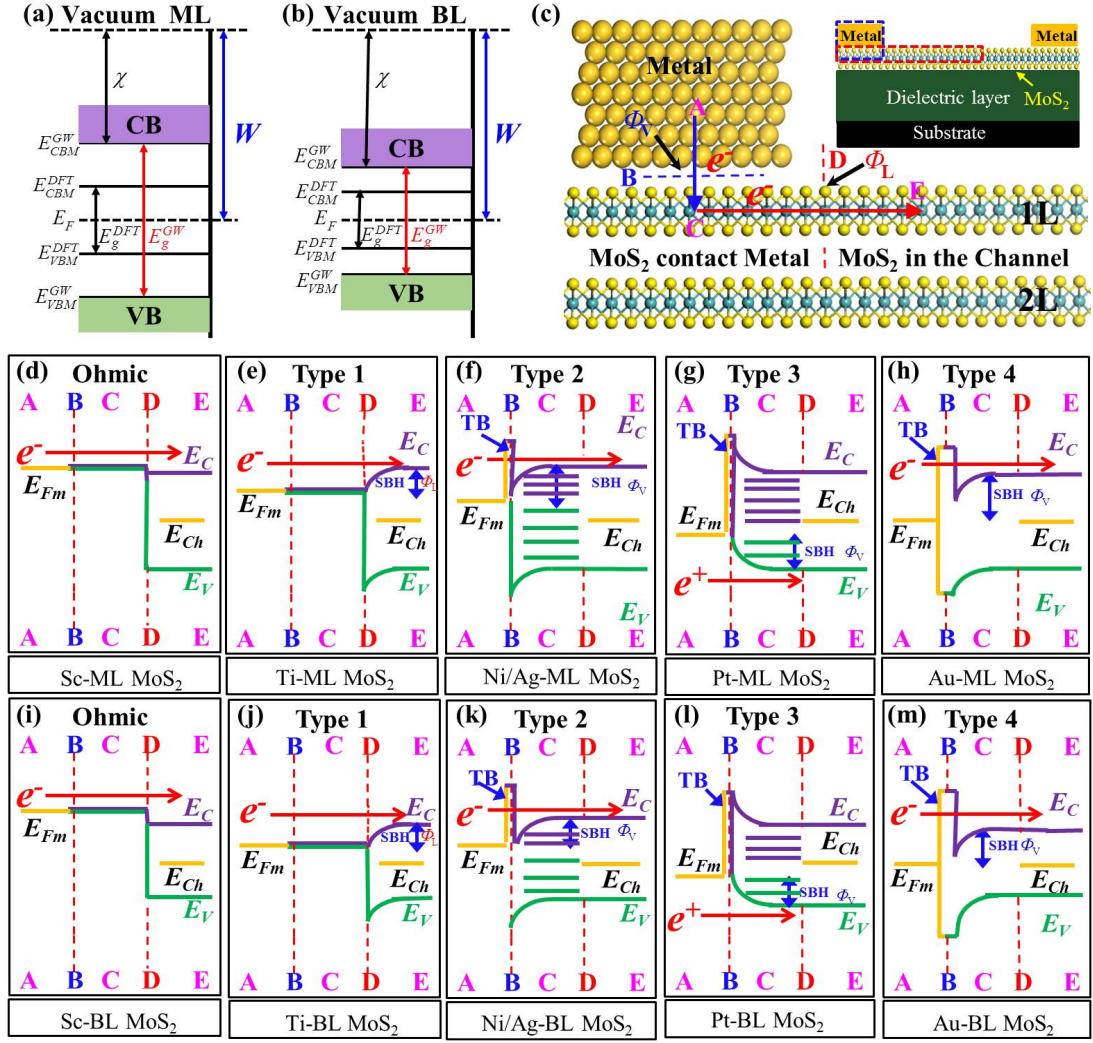

**FIG. 7.** Schematic illustration of the absolute band positions with respect to the vacuum level at both DFT and *GW* levels for ML (a) and BL (b) MoS$_2$, respectively. (c) Schematic cross-sectional view of a typical metal contact to 2D MoS$_2$. A, C, and E denote the three regions while B and D are the two interfaces separating them. Blue and red arrows show the pathway (A→B→C→D→E) of electron injection from contact metal (A) to the MoS$_2$ channel (E). Inset figure shows the typical topology of a MoS$_2$ FET. (d)-(m) Ten band diagrams of (c) at the DFT level, depending on the type of metals and MoS$_2$ layer number. TB denotes the tunneling transmission barrier. Examples are provided at the bottom (top) of each diagram. $E_{Fm}$ and $E_{Ch}$ denote the Fermi level of the absorbed system and the band gap center of channel MoS$_2$, respectively. Red arrows indicate the direction of electron or hole flow. The cause of the band bending is given in the main text.

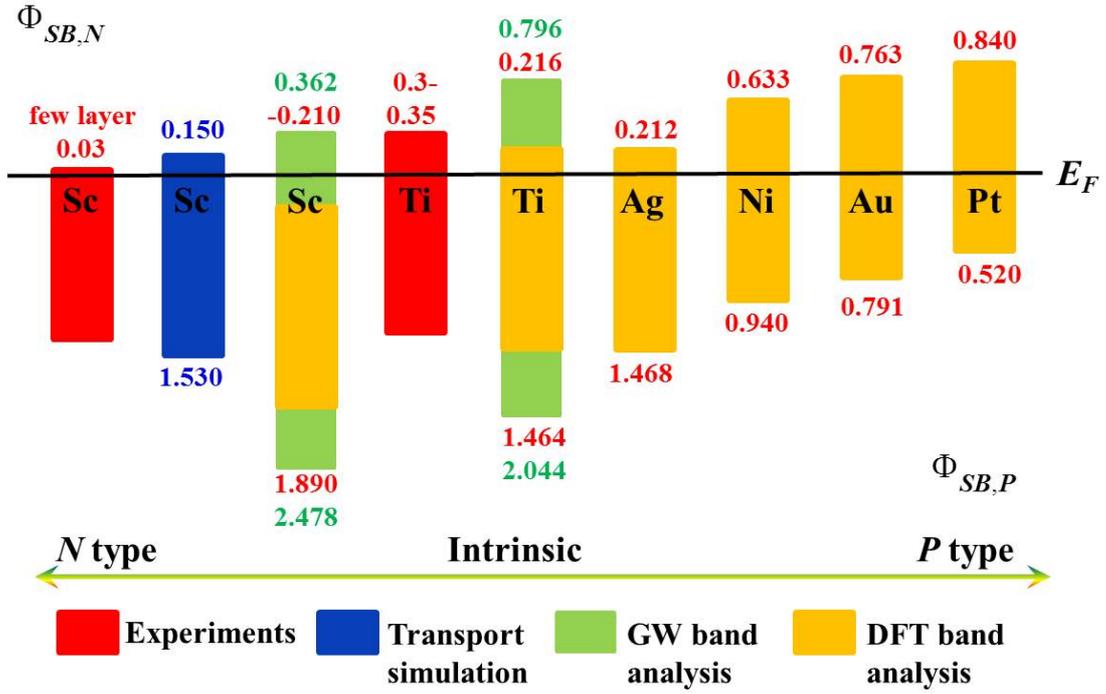

**FIG. 8.** $\Phi_{SB}$ of ML MoS$_2$ on the six metal surfaces. $\Phi_{SB,N}$ denotes $n$ type SB for electrons, while $\Phi_{SB,P}$ represents $p$ type SB for holes.

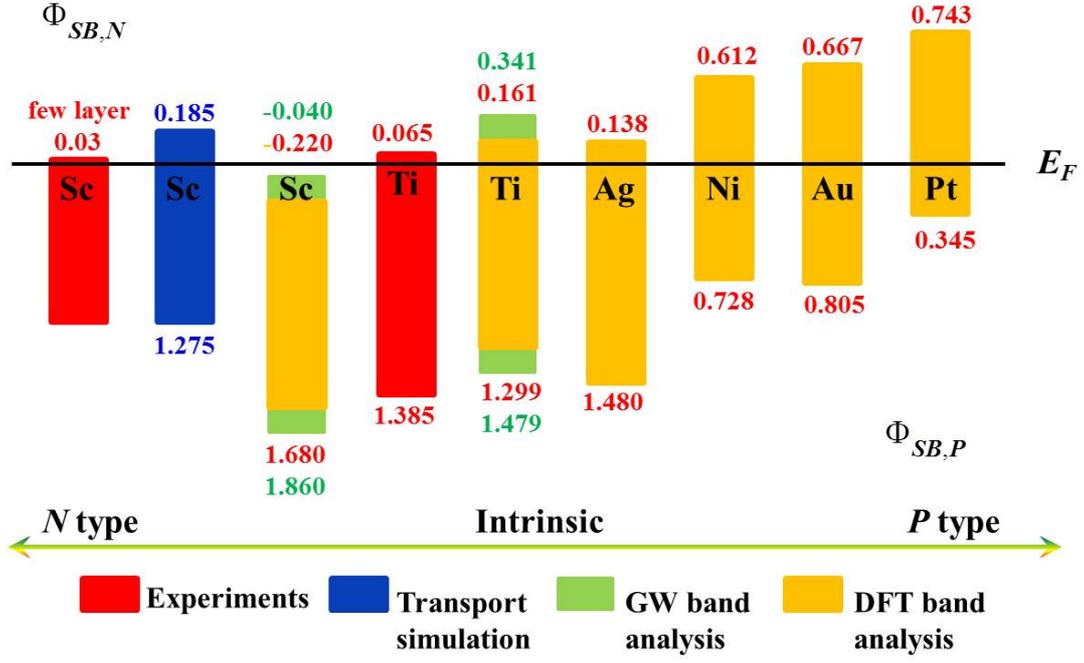

**FIG. 9.** $\Phi_{SB}$ of BL MoS$_2$ on metal surfaces. $\Phi_{SB,N}$ denotes *n* type SB for electrons, while $\Phi_{SB,P}$ represents *p* type SB for holes.

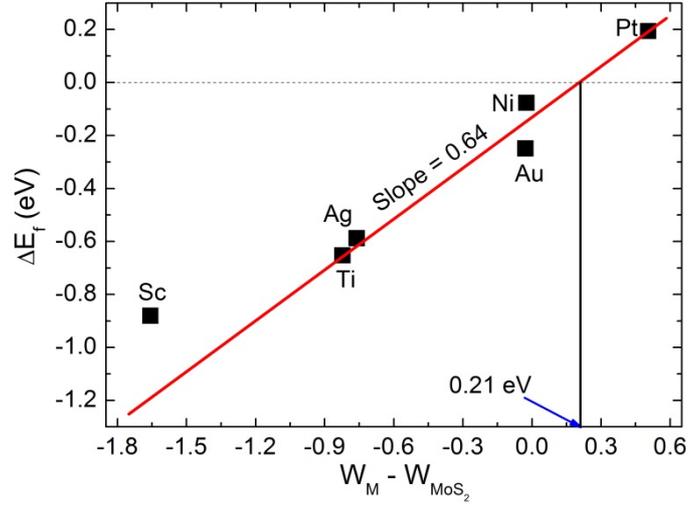

**FIG. 10.** Calculated Fermi level shift Δ$E_f$ as a function of $W_M - W_{MoS_2}$, the difference between the clean metal and ML MoS$_2$ work functions at the DFT level. $W_M - W_{MoS_2}$ = 0.21 eV is the cross point from *n*- to *p*-type doping. The red line is the fitting curve to the calculated points, and the slope of the line is 0.64.

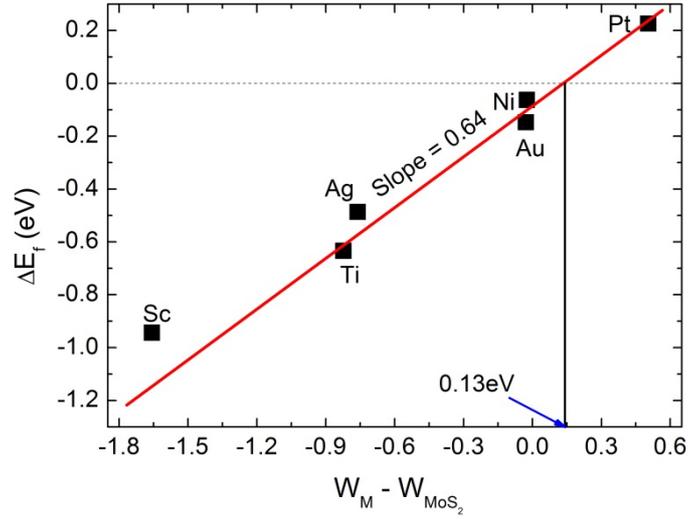

**FIG. 11.** Calculated Fermi level shift $\Delta E_f$ as a function of $W_M - W_{MoS_2}$, the difference between the clean metal and BL MoS$_2$ work functions at the DFT level. $W_M - W_{MoS_2} = 0.13$ eV is the cross point from *n*- to *p*-type doping. The red line is the fitting curve to the calculated points, and the slope of the line is 0.64.

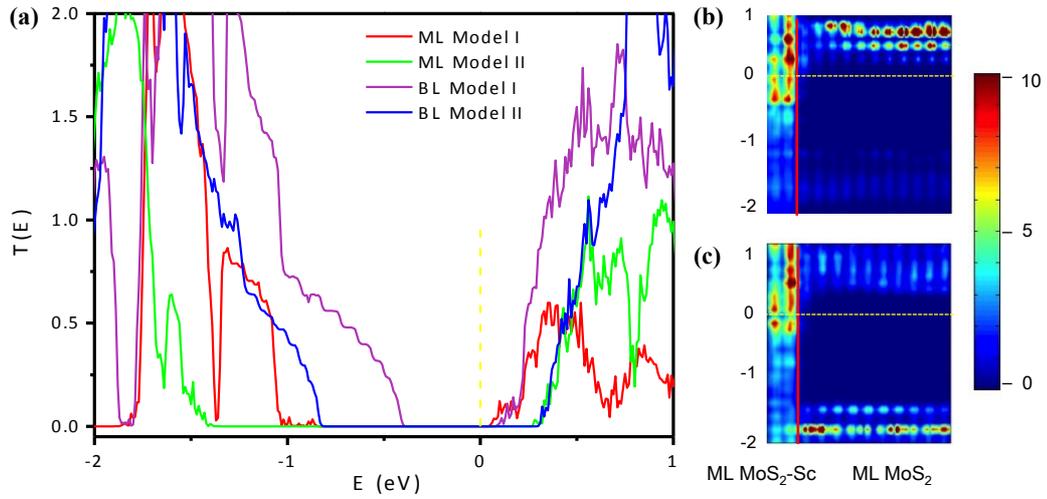

**FIG. 12.** (a) Transmission spectra of the ML and BL $MoS_2$ transistors with Sc electrodes. In Model I, the lattice constant of $MoS_2$ is adjusted to that of Sc, while in Model II the lattice constant of Sc is adjusted to that of $MoS_2$. (b-c) Local density of states (LDOS) in color coding for the ML $MoS_2$ transistors in Models I and II, respectively. The red line indicates the boundary of ML $MoS_2$-Sc and the free-standing ML $MoS_2$, and the yellow dashed line indicates the Fermi level.

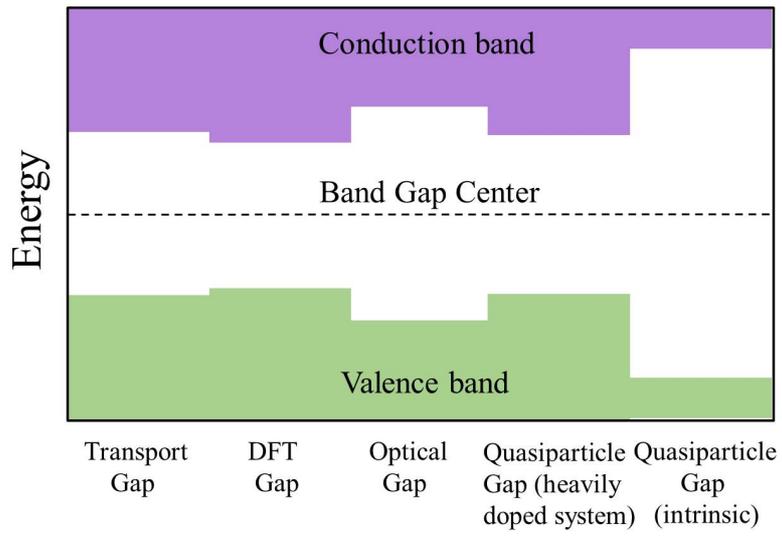

**FIG. 13.** A schematic diagram for the size relation of the five common band gaps of a 2D semiconductor.